# The effect of isovalent doping on the electronic band structure of group IV semiconductors

Maciej P. Polak*

*Department of Materials Science and Engineering, University of Wisconsin-Madison,
1509 University Ave., Madison, WI 53706, United States and
Department of Semiconductor Materials Engineering, Faculty of Fundamental Problems of Technology,
Wrocław University of Science and Technology, Wybrzeże Wyspiańskiego 27, 50-370 Wrocław, Poland*

Paweł Scharoch and Robert Kudrawiec[†]

*Department of Semiconductor Materials Engineering, Faculty of Fundamental Problems of Technology,
Wrocław University of Science and Technology, Wybrzeże Wyspiańskiego 27, 50-370 Wrocław, Poland*

Band gap engineering of group IV semiconductor has not been well explored theoretically and experimentally, except for SiGe. Recently, GeSn attracted a lot of attention due to a possibility of obtaining a direct band gap in this alloy, thereby making it suitable for light emitters. Other group IV alloys may also potentially exhibit material properties useful for device applications, expanding the space for band gap engineering in group IV. In this work the electronic band structure of all group IV semiconductor alloys is investigated. Twelve possible A:B alloys, where A is a semiconducting host (A = C, Si, and Ge) and B is an isovalent dopant (B = C, Si, Ge, Sn, and Pb), were studied in the dilute regime (0.8%) of the isovalent dopant in the entire Brillouin zone, and the chemical trends in the evolution of their electronic band structure were carefully analyzed. Density functional theory with state-of-the-art methods such as meta-GGA functionals and spectral weight approach to band unfolding from large supercells was used to obtain dopant-related changes in the band structure, in particular the direct band gap at the $\Gamma$ point and indirect band gaps at the $L(X)$ points of the Brillouin zone. Analysis of contributions from geometry distortion and electronic interaction was also performed. Moreover, the obtained results are discussed in the context of obtaining a direct fundamental gap in Ge:B (B = C, Sn, and Pb) alloys, and intermediate band formation in C:B (B = Sn and Pb) and Ge:C. An increase in localization effects is also observed: a strong hole localization for alloys diluted with dopant of larger covalent radius and a strong electron localization for alloys with dopant of smaller radius. Finally, it is shown that alloying Si and Ge with other elements from group IV is a promising way to enhance the functionality of group IV semiconductors.

## I. INTRODUCTION

Group IV semiconductors (Si and Ge) are fundamental components of Si-based electronics where semiconductor devices are grown on a Si substrate [1–4]. For low dimensional heterostructures (e.g., quantum wells or quantum dots) which are the core of current devices, band gap engineering is necessary. It is most often achieved by alloying materials originating from the same group [5].

In IV group systems it is possible to mix Si with Ge in the full composition range and obtain a SiGe alloy with good structural, electrical and optical properties [1–5]. The parent materials are in the same crystal structure (diamond) and have similar lattice constants, electronegativities and ionization energies, factors important for easy alloying of semiconductors. The SiGe alloy, however, has its own limitations. Throughout the entire composition range it is an indirect band gap material which limits its application in optoelectronics. It also only covers the span of energy gaps between that of Si and Ge. Alloying Si and Ge with other compounds from group IV (C, Sn, and Pb) is more challenging and, therefore, less explored experimentally [5]. However, it opens a way for further band gap engineering of group IV semiconductors. It may not only provide a possibility to obtain a group IV alloy with direct band gap but also may allow to cover a wider range of band gaps and lattice parameters. Moreover it is also interesting from a more fundamental point of view since C (diamond) is a wide gap material while Sn and Pb are metals.

Alloying Ge with Sn has been intensively explored experimentally in recent years, because it allows to achieve a material with direct band gap and to develop photonics integrated with the Si-platform [6–12]. On the other hand, GeSn with around 25% Sn is a zero band gap semiconductor/semimetal [13], which is interesting from the viewpoint of new phenomena in solid state physics [14, 15]. Research suggests that alloying Ge with Sn in a broad composition range is possible with present technologies, i.e., molecular beam epitaxy or chemical vapor deposition [16–19]. In other cases, where the size mismatch between the host and dopant atoms is more significant (e.g. CSn or SiPb) alloying in the full composition range can be more challenging or even impossible. However, in all cases alloying in a low concentration regime is most certainly possible by ion implantation and post implantation annealing [20, 21]. Despite this fact, in many cases this issue has still not be experimentally explored, i.e. the electronic band structure of most group IV alloys is still unknown. Therefore, the results of theoretical investigations such as the one presented here may motivate experimental researchers to pursue their synthesis and experimental characterization.

From the point of view of Si-based technology, Si- and Ge-based alloys with low concentration of C, Sn, and Pb are the most interesting since it is important to tune the conduction and valence bands without drastic modification of the lattice constant. Such conditions are typical for highly mismatched alloys (HMA) i.e., alloys composed of atoms with significant differences in valence radii, electronegativities and ionization energies [22–26].

---

* maciej.polak@pwr.edu.pl
† robert.kudrawiec@pwr.edu.pl

Among group IV alloys such conditions may be present in many systems but the electronic band structure has not been extensively analyzed. Just a handful of published theoretical and experimental papers only for selected systems [27–39] and no cross-sectional work can be found for such alloys. A systematic study and comparison of the group IV alloys would provide a valuable insight into the chemical trends, which this paper aims to provide and which have not been studied previously.

In this work we present systematic and comprehensive first-principle density functional theory (DFT) studies of the effect of isovalent doping on the electronic band structure of group IV semiconductors. The obtained results are analyzed in terms of the influence of the optimization of the geometry (the distortion of atomic positions from pure material equilibrium) and the influence of the element substitution on the changes in the electronic band structure. The observed changes in the electronic band structure are discussed in the context of band anticrossing model, useful for simplified theoretical modeling of HMAs.

## II. RESULTS AND DISCUSSION

To graphically show the mismatch between alloying elements, differences of covalent radii and 1st ionization energies between are presented in Fig. 1. Throughout the text a naming convention of A:B has been adopted, where A indicates the host material, while B stands for the isovalent dopant, i.e. a low composition alloy/solid solution. In Fig. 1, the studied alloys grouped in terms of the mismatch: a) HMAs where isovalent dopant is smaller and has higher ionization energy (significant changes in the conduction band), a*) subgroup of a) where the magnitude of mismatch causes an emergence of additional (conduction) band, b) similar alloying elements (regular alloys), c) HMAs where isovalent dopant is larger and has lower ionization energy (significant changes in the valence band), c*) subgroup of c) where the magnitude of mismatch causes an emergence of additional (valence) band.

From the materials in a) and c) groups a different composition dependence of the band gap is expected than that in the traditional semiconductor alloys b), which will be discussed more in depth further in the text. The estimated band gap reductions (per % of the composition) were collected in Tab. I. The reduction of the gap is always higher than that expected from the virtual crystal approximation between host materials, indicating a stronger than linear change with composition. It can be described in various ways, depending on the type of alloying elements (Eqs. 1, 2 and 3 discussed later). Therefore, the parameters from Tab. I should be considered as an approximation in the dilute regime (composition of up to a few %). In all of the cases, except Si:C and Ge:Si, the band gap of the material is reduced due to alloying.

Figure 2 contains the unfolded band structures of all the studied alloys, C-based, Si-based and Ge-based, respectively,

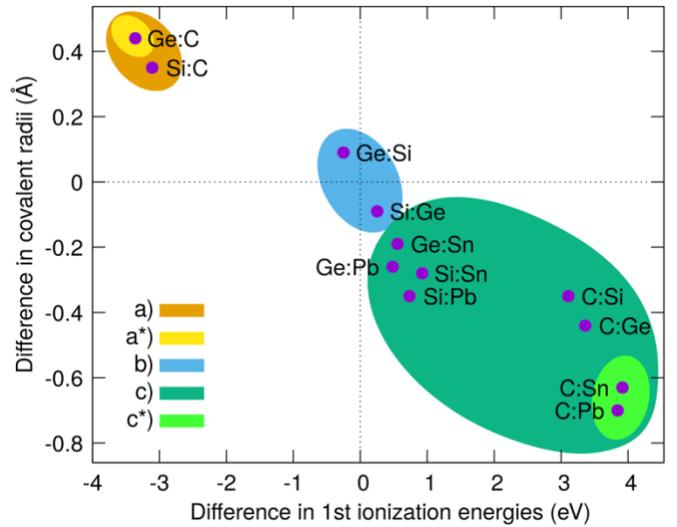

FIG. 1: Differences in covalent radii [40] and 1st ionization energies [41] between the alloying elements as a representation of atomic mismatch. a) HMAs where isovalent dopant is smaller and has higher ionization energy, a*) subgroup of a) where the magnitude of mismatch causes an emergence of additional (conduction) band, b) similar alloying elements (regular alloys), c) HMAs where isovalent dopant is larger and has lower ionization energy, c*) subgroup of c) where the magnitude of mismatch causes an emergence of additional (valence) band.

TABLE I: Total change of the direct and indirect band gaps per 1% of isovalent dopant. The columns are arranged in a similar fashion to Fig. 2 with rows representing the host material, and the columns representing the dopant. The indirect gap corresponds to the X valley in BZ for C and Si and the L point for Ge.

| Direct (eV/1%) | C | Si | Ge | Sn | Pb |
|---|---|---|---|---|---|
| C | 0 | -0.134 | -0.307 | -1.022 | -1.541 |
| Si | 0.019 | 0 | -0.003 | -0.024 | -0.117 |
| Ge | -0.593/0.140 | 0.024 | 0 | -0.045 | -0.099 |
| Indirect (eV/1%) | C | Si | Ge | Sn | Pb |
| C | 0 | -0.158 | -0.299 | -0.961 | -1.441 |
| Si | -0.016 | 0 | -0.002 | -0.026 | -0.075 |
| Ge | -0.011 | 0.003 | 0 | -0.030 | -0.066 |

with other group IV (C, Si, Ge, Sn and Pb) isovalent elements as dopants in $1/128 \approx 0.8$ % concentration. The logic of the rows and columns is arranged as follows: each row represents one host material, C, Si, and Ge, top to bottom, and each column represents an isovalent dopant, C, Si, Ge, Sn and Pb, left to right. Therefore, panels a), g), and m) correspond to pure host materials. For comparison, the material parameters for group IV materials involved here (including host materials and remaining dopants) are gathered in Tab. II.

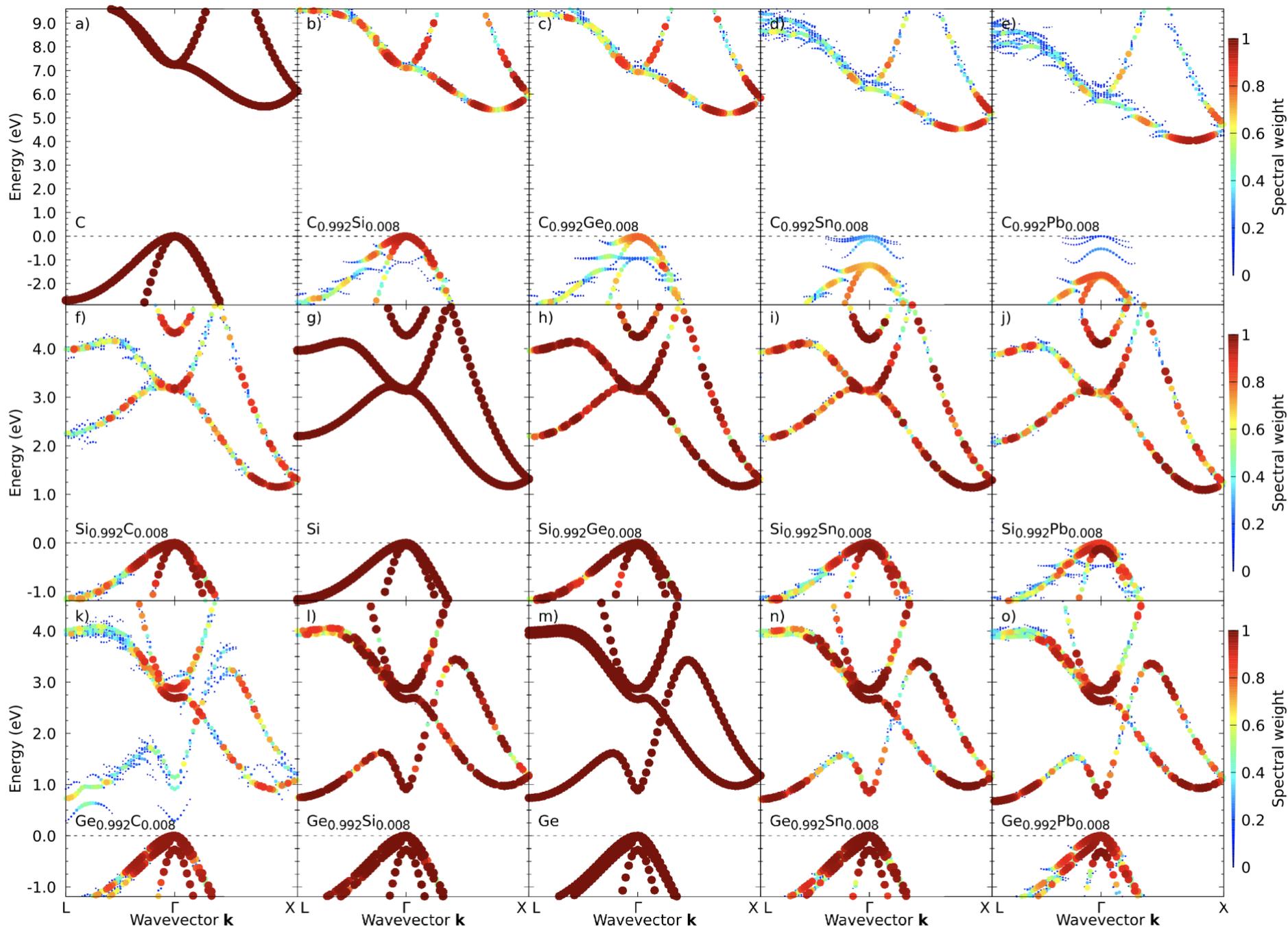

FIG. 2: Unfolded band structures of all the studied alloys organized row- and column-wise. Host material changes from top to bottom with increasing atomic number (C, Si, and Ge) while isovalent dopant changes from left to right with increasing atomic number (C, Si, Ge, Sn, and Pb).

TABLE II: Material parameters obtained from calculations. See Sec. III for more details.

|  | C | Si | Ge | Sn | Pb |
|---|---|---|---|---|---|
| $a_0$ (Å) | 3.536 | 5.404 | 5.648 | 6.481 | 6.852 |
| $E_g^\Gamma$ (eV) | 7.4 | 3.2 | 0.89 | -0.39 | -4.633 |
| $E_g^L$ (eV) | 9.55 | 2.25 | 0.74 | 0.086 | -1.99 |
| $E_g^X$ (eV) | 5.57 | 1.21 | 0.97 | 0.711 | 0.028 |
| $c_{MBJ}$ | 1.463* | 1.114* | 1.210* | 1.197 | 1.212 |

The 0.8 % composition is obtained by replacing one of the atoms in a 128-atom supercell with an isovalent dopant. Using a single composition allows to draw more meaningful conclusions about the chemical trends present within the entire family of the studied systems and direct comparison between them. The dilute regime is also most likely to be most relevant in real life application, where the highly mismatched nature of the systems will only allow for the synthesis of low composition materials. The qualitative result of different band structures in Fig. 2 can be categorized in three groups: band structure almost unchanged (except for the change in the band gap) as in regular semiconductor alloys (Si:Ge and Ge:Si (Fig. 2 h) and l)), significant changes in the Bloch character (spectral weights: color and size of the points) and disorder in the band structure mostly in the valence band (C:Si, C:Ge, C:Sn, S:Pb, Si:Sn, Si:Pb, Ge:Sn and Ge:Pb (Fig. 2 b), c), d), e), i), j), n) and o)), and similarly pronounced changes but mostly in the conduction band (Si:C and Ge:C (Fig. 2 f) and k)). Among the latter two categories, in particular the cases of C:Sn, C:Pb and Ge:C (Fig. 2 d), e) and k)) should be highlighted as those where distinct new bands appear. This categorization can be explained in terms of mismatch between the host and isovalent dopant as seen in Fig. 1.

From Tab. I it can be predicted that in some of the alloys, a possibility of a transition from an indirect to direct band gap exists. In order for this to happen the reduction of the indirect gap has to be lower than that of the direct gap, and the band gap has to wide enough to allow for the change in the band gap character, while the gap is still present. Data in Tabs. I and II allow to estimate that in line with previous studies, this certainly occurs for Ge:Sn, and seem to be also possible for Ge:Pb. A particularly interesting behavior can be found for Ge:C, where a direct band gap is achieved by an additional band emerging inside the host Ge band gap due to the introduction of carbon (composed of carbon *s* orbitals). The other materials, although most do exhibit a higher reduction in the direct than indirect band gap, cannot accommodate such a transition for a composition low enough to keep the band gap open.

Apart from the very apparent conclusion about the chemical trends in the observed changes in the band structure, where the magnitude of the band gap reduction increases with the mismatch of the host and impurity atoms, the three specific cases of Ge:C, C:Sn and C:Pb are particularly interesting in terms of their unique band structure. The intermediate conduction band appearing in Ge:C, even for small concentrations of C, such as the 0.8% presented here, drastically changes the character of the material. Although both the host Ge and the impurity C are indirect band gap semiconductors, the influence of C atoms in Ge host result in a direct gap material, suitable for light emission. In addition, the small admixture of C does not strongly influence the lattice constant of the Ge host enabling growth Ge:C alloy on Ge or Si substrates. Because of this, Ge:C can be considered as a good candidate for light emission within the group IV semiconductor devices. Previously, a very similar change in the electronic band structure was observed for GaP diluted with nitrogen [42], where the indirect GaP host, after introducing even small amounts of nitrogen, exhibits photoluminescence [43, 44]. The similarity of the Ge:C suggests that even lower concentrations of C than the 0.8% studied here would allow to achieve this effect.

The other two intriguing band structures of C:Sn and C:Pb do not seem to be promising for light emitters because of the indirect gap. Although clearly separate valence bands appear, which are promising in intermediate band solar cells [45], the very wide band gap of the host diamond strongly limit its application as a light absorber in solar cells. However, these alloys can be useful in another applications and they are very interesting themselves for better understanding the solid state physics. For example, the presence of two valence sub-bands separated by certain energy gap should result in a specific spectral dependence of absorption. The presence of the intermediate band in this case can help in p-type doping of diamond.

Out of the 12 diluted alloys studied here, two (Si:Ge and Ge:Si, Fig. 1 a)) are well known as regular, well matched semiconductor systems. The composition dependence of band gaps in these systems is well described by the quadratic Vegard's law:

$$E_g^{A_{1-x}B_x}(x) = (1-x)E_g^A + xE_g^B - bx(1-x) \qquad (1)$$

where $E_g^A$ and $E_g^B$ are the band gaps of the host materials A and B, respectively, *x* is the composition, and *b* is a bowing parameter.

C:Si and Si:C are less similar to regular alloys since the spectral weights of wave functions significantly decrease for both the conduction and the valence band, see Fig. 2. In addition, the dopant-related change in the band gap is much larger than that observed for regular alloys, which in case of group IV alloys are represented by SiGe, see Tab. I.

Very similar changes in the electronic band structure are observed for Ge:Sn. From our previous research [13] as well as from literature [23] it is known that Ge:Sn has a different behavior than regular alloys, which can be attributed to the higher mismatch in size and electronegativity between Ge and Sn atoms, allowing this material to be classified as a HMA. A valence band anticrossing model (VBAC) has been shown to accurately reproduce the composition dependence of band gap in this alloy [23] in the low concentration region. The composition dependent band gap in the VBAC model is expressed by replacing the $bx(1-x)$ term in Eq. 1 with a two-parameter expression, dependent on the relative position of the impurity B atom resonant level to the valence band maximum (VBM),

$E_{BI}$, and a parameter $C_{BM}$ describing the magnitude of the interaction [46]. The equation then reads:

$$E_g^{A_{1-x}B_x}(x) = (1-x)E_g^A + xE_g^B - \frac{1}{2}\left(\sqrt{E_{BI}^2 + 4C_{BM}^2 x} - E_{BI}\right) \quad (2)$$

Since the magnitude of the interaction increases with the mismatch between the alloying elements [47–49], it is reasonable to expect similar behavior of seven other systems studied here, i.e. Ge:Pb, Si:Sn, Si:Pb, C:Si, C:Ge, C:Sn and C:Pb (Fig. 1 c)). From those, particularily interesting are C:Sn and C:Pb, where due to the extreme mismatch (Fig. 1 c*)), a separate valence band emerges.

The remaining Ge:C alloy exhibits an opposite behavior, similar to the one observed in dilute nitrides such as GaP:N or GaAs:N [42], where the anticrossing takes place in the conduction band due to an opposite mismatch between the Ge and C atoms as compared to the mismatch between Ge and Sn/Pb. In this case the conduction band is split in two unfilled subbands, $E_-$ and $E_+$, according to the (conduction) band anticrossing model (BAC) [22]:

$$E_\pm^{A_{1-x}B_x}(x) = \frac{1}{2}\left(E_g^A + E_{BI} \pm \sqrt{(E_g^A - E_{BI})^2 + 4C_{BM}^2 x}\right) \quad (3)$$

where, similarly to the VBAC case, $E_{BI}$ is the energetic distance form the impurity atom resonant state to the VBM and $C_{BM}$ is the magnitude of the interaction. It is worth noting that this equation, unlike Eq.2, is only applicable in a regime of relatively low concentration of impurities. This case is shown in Fig. 1 a*) and in Fig. 2 k).

The BAC parameters can be determined experimentally by studying the pressure dependence of energies of $E_-$ and $E_+$ transitions [22] or the content dependence of these transitions [26]. In general parameters for VBAC can be also obtained by experimental measurements of $E_-$ and $E_+$ transitions, but the valence band is more complex than the conduction band and to date the $E_-$ transition has not been clearly observed for any alloy described within the VBAC. For the same reason, a direct accurate extraction of the valence band anticrossing parameters is impossible from our calculations (i.e., three valence bands interacting with the impurity levels lead to 6×6 Hamiltonian [23, 50] and, therefore, complex equations). However, a less rigorous ballpark estimate can be inferred for the BAC in the case of Ge:C. The two subbands observed in the band structures, corresponding to $E_-$ and $E_+$ in Eq. 3, allow to directly solve Eq. 3 for $E_{BI}$ and $C_{MB}$ at $x = \frac{1}{128}$. The solution yielded $E_{BI} \approx 0.43$ eV and $C_{MB} \approx 3.12$ eV.

In the case of VBAC, however, even though values cannot be directly extracted, the analysis of the band structures in Fig. 2 allows to draw conclusion on the expected behavior of $E_{BI}$ and $C_{BM}$ parameters. The impurity level should be closer to the valence band edge as the radius of the isovalent dopant increases and ionization energy decreases. Similarly, the interaction energy should increase. This is most easily noticeable in Fig. 2 b)-e). The visual analysis of the unfolded band structures in Fig. 2 allows for another qualitative conclusion. The Bloch spectral weight, is connected to the localization of states, which increases with the decrease in spectral weight [51]. Here, again, the loss of spectral weights follows the trends of mismatch (Fig. 1), with the most disorder and highest loss of Bloch character for the most mismatched alloys.

To gain a better insight into the origin of the observed changes in the electronic band structure, additional calculations were performed, where the analysis of alloys was separated into two contributions: the influence of the distortion in geometry (the ion coordinates) and the influence of the elements substitution. The first one was obtained by using the crystal structure optimized for the alloy, but keeping the material pure by populating the atomic sites with only the host material atoms. Second, the crystal structure has been kept undisturbed, in the equilibrium diamond structure ionic positions, but the atoms have been replaced to form an alloy. This approach allowed to decompose the geometric and atomic contributions to the changes of the electronic band structure and its features. The results in terms of the change in band gap per one % of the isovalent dopant are gathered in Tabs. III.

Figure 3 presents an example of the unfolded band structures for the three cases: all effects taken into account (geometry optimization and element substitution) (Fig. 3d), only the optimized lattice geometry distortion included (Fig. 3b), only the electronic effects included (element substitution) (Fig. 3c), and a pure material for comparison (Fig. 3a). The band gap reduction decomposition for the remaining 11 materials presented in a similar fashion to this in Fig. 3 are gathered in the supplementary material as Figs. S1-S11. From the analysis of this data it is clearly visible that the change of the band structure can be attributed to both the lattice geometry distortion and the electronic contribution of the introduced atom. Often the magnitude of the two contributions is comparable. Table III shows that as the mismatch between the atoms increases, the contribution of the local lattice distortion plays increasingly important role in the band structure change. This indicates that virtual crystal-like approximations such as the alchemical mixing method [52], unable to account for the changes in geometry, should be avoided. This is in addition to the already mentioned strongly nonlinear overall band gap reduction.

The presented studies of the electronic band structure of group IV diluted semiconductors have been discussed in the context of their potential applications in light emitters or solar cells. However, the most interesting seems to be the discussion of changes in the electronic band structure in the context of chemical trends. For the studied alloys the electronic band structure is changing mainly in the valence band when small atoms are replaced by large atoms, see all C:B alloys, where B is the isovalent dopant, Si:Sn and Si:Pb. An opposite behavior is observed when large atoms are replaced by small atoms, see Ge:C, which is the opposite situation to C:Ge. For all these alloys the spectral weight of wavefunction strongly decreases for the modified band. It means a strong hole (electron) localization for IV alloy diluted with dopant of larger (smaller) valence radius. For remaining alloys changes in the electronic band structure are comparable for both the conduc-tion and the valence band. For Si:Ge and Ge:Si these changes are very small since differences in electronegativities, ionization energies and atomic radii are very small for Si and Ge.

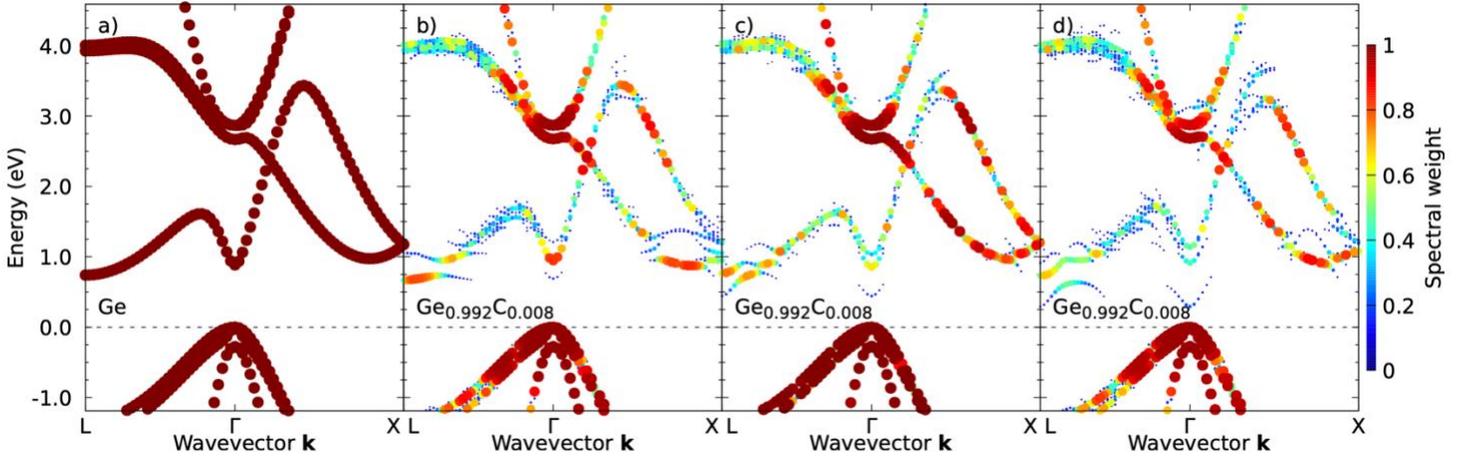

FIG. 3: Unfolded band structures of Ge:C showing the contribution of geometry distortion (b) and electronic (c) to band gap change, together with the band structure of the Ge:C alloy (d) and pure Ge (a).

TABLE III: Decomposition of direct band gap change into geometric and electronic contributions. The indirect gap corresponds to the X valley in BZ for C and Si and the L point for Ge.

| Material | Change of band gap per % (eV) (Geometry distortion contribution) | | Change of band gap per % (eV) (Electronic contribution) | |
|---|---|---|---|---|
| | direct | indirect | direct | indirect |
| C:Si | -0.082 | -0.094 | -0.029 | -0.089 |
| C:Ge | -0.122 | -0.161 | -0.143 | -0.153 |
| C:Sn | -0.272 | -0.439 | -0.924 | -0.826 |
| C:Pb | -0.394 | -0.691 | -1.724 | -1.552 |
| Si:C | 0.016 | -0.051 | 0.006 | 0.003 |
| Si:Ge | 0.001 | 0.004 | -0.003 | -0.003 |
| Si:Sn | 0.002 | 0.000 | -0.018 | -0.016 |
| Si:Pb | 0.002 | -0.005 | 0.008 | -0.033 |
| Ge:C | 0.061 | -0.073 | 0.059 | 0.026 |
| Ge:Si | 0.006 | 0.000 | 0.014 | 0.003 |
| Ge:Sn | -0.037 | -0.016 | 0.009 | -0.014 |
| Ge:Pb | -0.057 | -0.029 | -0.022 | -0.024 |

In general similar chemical trends can be expected for group III-V and II-VI semiconductors, but these material system are more complex to investigate and interpret because of the anion and cation sublattice. Group IV family of alloys are the simplest systems that vary from a semiconductors with a wide gap (diamond) to metals (lead) and is therefore interesting to consider, even if some of these alloys may be difficult to synthesize.

### III. METHODS

Alloys were modeled using a 128-atom supercell (4x4x4 multiplication of a primitive 2-atom unit cell). In each supercell one host atom was replaced with the alloying element in order to obtain a dilute concentration of around 0.8%. The LDA exchange-correlation functional [53] was used in the geometry optimization, since for the studied group-IV materials it has been proven to outperform other functionals [54]. For the alloys, the internal atom positions were fully optimized for a fixed lattice constant given as a linear interpolation between the pure hosts material in their diamond crystal structure.

The electronic band structure was calculated with the mBJLDA functional [55] to overcome the problem of band gap underestimation in LDA. Even though the mBJLDA functional reproduces the band gap of group IV semiconductors with a relatively high accuracy [55, 56], the $c$ parameter for host materials was slightly increased (by no more than 5% from the self-consistently calculated value, see Tab. II) to achieve a perfect agreement with the 0 K values that are well established from experimental measurements [5]. For Sn and Pb, since their band structure in diamond crystal lattice is not experimentally known, it has been calculated self consistently. The $c$ parameter used to calculate the electronic band structure has been obtained as a linear interpolation of the values for parent materials. This approach has been thoroughly tested and proven to be very successful in our previous work [13] as well as in other research [57].

The lattice constants obtained for the pure host materials, as well as the self-consistent and adjusted values of the $c$ parameters, together with the obtained band gap values are presented in Tab. II. The supercell approach produces a band structure obfuscated by the band folding due to the reduction of the Brillouin zone (BZ). To recover the full primitive BZ band dispersion, the spectral weight approach to band unfolding was used [58, 59], by the means of the `fold2bloch` code [60].

Convergence studies of the control parameters were performed and as a result, a 2 ×2 ×2 Monkhorst-Pack k-point mesh [61] (an equivalent of an 8 × 8 × 8 mesh in the primitive unit cell) was used, with a $10^{-3}$ eV/Å tolerance on the maximum residual force in the geometry optimization and $10^{-6}$ eV total energy convergence in the electronic band structure calculations.

Spin-orbit interactions were included in all calculations. The recommended set of PAW potentials [62] were used, with the $s$ and $p$ states treated as valence in C, Si and Ge, and additionally taking into account $d$ states in the heavier Sn and Pb. All DFT calculations were performed using the VASP package [63, 64].

## IV. SUMMARY

The effect of isovalent doping on the electronic band structure of group IV semiconductors has been studied within the density functional theory with state-of-the art methods including the meta-GGA mBJLDA functional and band unfolding based on the spectral weight approach. It has been found that the dopant-related changes in the band gap (both direct and indirect) occur according to chemical trends in differences (mismatch) in atom radii and ionization energies. The changes in band structure are accompanied by an increase in localization of electronic states, indicated by a loss of Bloch character of conduction and valence bands. Moreover it has been concluded that a direct gap can be achieved in Ge via alloying with C, Sn or Pb. In the case of Ge:C the direct band gap is achieved by emergence of an intermediate empty band below the conduction band minimum. In the two other cases, Ge:Sn and Ge:Pb, the direct band gap is achieved due to faster rate of reduction of the direct band gap than the indirect band gap as a function of composition. The same effects are most likely impossible to achieve in C and Si by alloying with low compositions of other group IV elements since the electronic band structure of the host material is strongly indirect, i.e., the energy difference between the direct and indirect gap is much larger than the dopant-related changes in band gaps. The tuning of band gap from indirect to direct in Ge strongly enhances the functionality of group IV semiconductors since it can make it more useful for optoelectronics, i.e. the field that until now is dominated by III-V semiconductors. The influence of local lattice optimization has been investigated as well, revealing that it has a significant effect on the modification of the band structure, in particular in alloys with high mismatch, and should not be dismissed. Another important finding is the formation of an intermediate band, which has been clearly observed in the conduction band for Ge diluted with C and in the valence band for C diluted with Sn and Pb. This property can also expand the functionality of group IV semiconductors, since materials with intermediate bands may have perspectives for use in new generation of solar cells or other applications.

## V. SUPPLEMENTARY MATERIAL

Additional details on the performed calculations as well as supplementary figures are available as supplementary material attached to the online version of this paper.

## ACKNOWLEDGMENTS

This work has been partially funded within the grant of the National Science Center Poland (OPUS11, UMO-2016/21/B/ST7/01267). In addition M. P. Polak acknowledges the support within National Science Center Poland (ETIUDA no. 2016/20/T/ST3/00258). Calculations have been carried out using resources provided by Wroclaw Centre for Networking and Supercomputing.